\documentclass{tise_style}\usepackage[]{graphicx}\usepackage[]{color}
\makeatletter
\def\maxwidth{ %
  \ifdim\Gin@nat@width>\linewidth
    \linewidth
  \else
    \Gin@nat@width
  \fi
}
\makeatother

\definecolor{fgcolor}{rgb}{0.345, 0.345, 0.345}
\newcommand{\hlnum}[1]{\textcolor[rgb]{0.686,0.059,0.569}{#1}}%
\newcommand{\hlstr}[1]{\textcolor[rgb]{0.192,0.494,0.8}{#1}}%
\newcommand{\hlcom}[1]{\textcolor[rgb]{0.678,0.584,0.686}{\textit{#1}}}%
\newcommand{\hlopt}[1]{\textcolor[rgb]{0,0,0}{#1}}%
\newcommand{\hlstd}[1]{\textcolor[rgb]{0.345,0.345,0.345}{#1}}%
\newcommand{\hlkwc}[1]{\textcolor[rgb]{0.333,0.667,0.333}{#1}}%
\newcommand{\hlkwd}[1]{\textcolor[rgb]{0.737,0.353,0.396}{\textbf{#1}}}%

\usepackage{framed}
\makeatletter
\newenvironment{kframe}{%
 \def\at@end@of@kframe{}%
 \ifinner\ifhmode%
  \def\at@end@of@kframe{\end{minipage}}%
  \begin{minipage}{\columnwidth}%
 \fi\fi%
 \def\FrameCommand##1{\hskip\@totalleftmargin \hskip-\fboxsep
 \colorbox{shadecolor}{##1}\hskip-\fboxsep
     \hskip-\linewidth \hskip-\@totalleftmargin \hskip\columnwidth}%
 \MakeFramed {\advance\hsize-\width
   \@totalleftmargin\z@ \linewidth\hsize
   \@setminipage}}%
 {\par\unskip\endMakeFramed%
 \at@end@of@kframe}
\makeatother

\definecolor{shadecolor}{rgb}{.97, .97, .97}
\definecolor{messagecolor}{rgb}{0, 0, 0}
\definecolor{warningcolor}{rgb}{1, 0, 1}
\definecolor{errorcolor}{rgb}{1, 0, 0}
\newenvironment{knitrout}{}{} 

\usepackage{alltt}
\usepackage[utf8]{inputenc}
\usepackage{graphicx}
\usepackage{amssymb}
\usepackage{amsmath, mathtools, bbm, bm, multicol, booktabs}
\usepackage{url}
\mathtoolsset{showonlyrefs}
\usepackage{subcaption}
\bibpunct{(}{)}{;}{a}{}{,}

\title{Data Visualization on Day One: Bringing Big Ideas into Intro Stats Early and Often
}

\author{Xiaofei Wang, Cynthia Rush, and Nicholas Jon Horton}
\IfFileExists{upquote.sty}{\usepackage{upquote}}{}
\begin{document}

\maketitle
\begin{center}
\today
\end{center}
\newpage

\begin{center}
\textbf{Abstract}

\end{center}

In a world awash with data, the ability to think and compute with data has become an important skill for students in many fields. For that reason, inclusion of some level of statistical computing in many introductory-level courses has grown more common in recent years. Existing literature has documented multiple success stories of teaching statistics with R, bolstered by the capabilities of R Markdown.
In this article, we present an in-class data visualization activity intended to expose students to R and R Markdown during the first week of an introductory statistics class. The activity begins with a brief lecture on exploratory data analysis in R. Students are then placed in small groups tasked with exploring a new dataset to produce three visualizations that describe particular insights that are not immediately obvious from the data. Upon completion, students will have produced a series of univariate and multivariate visualizations on a real dataset and practiced describing them.

\vspace*{.3in}


\newpage

\section{INTRODUCTION}
A number of
calls to infuse our statistics curricula with statistical computing to expose students to authentic data experiences have been recently published \citep{Nolan:2012bb, Hardin:2014vl, HortHard:2015}. In the revised 2016 Guidelines for Assessment and Instruction in Statistics Education (GAISE) College Report, instructors are recommended to ``integrate real data with a context and a purpose'', to ``use technology to explore concepts and analyze data'', and to ``emphasize the multivariate nature of the discipline"  \citep{GAISEcollege}. With booming interest in data science, it is now more important than ever to bring statistical software into the classroom early, to enable analysis of real-world data, to expose students to the excitement and potential of statistics, and to provide examples where insights are extracted from data.

\cite{Chance:2007vt} provides a broad overview of how technology benefits learning even at the introductory level, citing various tools ranging from graphing calculators to computing software. With the advent of R \citep{Team:wsDFV8Zj} and RStudio \citep{Team:AGFjvv8f}, students now have access to a free, open-source software with an established prominence in industry. A powerful software such as R does come with a steep learning curve for some; we believe that using data visualization as the entry point to learning R and introductory statistics could lessen the anxiety associated with both. To facilitate this initial foray into the world of data visualization using R, we advocate the use of (1) R Markdown \citep{RMarkdown}, a system that is integrated into RStudio, because its workflow inherently encourages well-documented reproducible analysis \citep{Baumer:2014ud} and (2)  the {\tt mosaic} package \citep{Pruim:2015ww}, whose modeling language provides a simplified interface to multivariate descriptive statistics, linear models, and graphical displays.

Much has been written about the nuances of data visualization techniques; see for example \cite{tufte1983visual, cleveland1985elements} and, more recently, \cite{wickham2009ggplot2, wickham2010layered}. \cite{Nolan:2015ua} describes the potential for visualization to inform statistical thinking and suggests ways to incorporate this capacity into statistics courses, noting that computational advancements in recent decades have made it far easier for students to apply advanced graphical tools at the level of introductory statistics.



In this article, we present an in-class introductory multivariate data visualization activity designed for a single class period during the first week of class. The activity begins with a brief, instructor-led introduction to exploratory data analysis in R. Students are then placed in small groups tasked with exploring a new dataset to produce three visualizations that describe particular insights that are not immediately obvious from the data. Upon completion, students will have produced a series of univariate and multivariate visualizations on a real dataset and practiced describing them.

\newpage

\section{THE ACTIVITY} 
\label{sec:the_activity}
\subsection{Overview} 
\label{sec:overview}
The proposed activity begins with a 15-minute tutorial led by the instructor on how to generate basic numeric summaries and visualizations in R. The lecture introduces the basic functionality used by the {\tt mosaic} R package to simplify the process of generating multivariate graphical displays and summary statistics. The focus of this tutorial is not on the mechanics of these exploratory tools (for example, how we compute the heights of a histogram) but rather on enhancing one's comprehension of the relationships between variables in a dataset. To that end, a complex, multivariate dataset should be selected to serve as the basis of this tutorial. In our case, we used the baseline data from the Health Evaluation and Linkage to Primary Care Clinical Trial (`HELPrct'), which enrolled subjects without primary medical care while they were attending a substance-use detoxification unit \citep{samet2003linking}. Table~\ref{tbl:helprct} summarizes a portion of the variables collected at baseline that might be of interest to study.

\begin{table}
  \caption{Abridged codebook for the `HELPrct' dataset}\label{tbl:helprct}
  \begin{tabular}{@{}lp{0.9\textwidth}@{}}
  \toprule
  Variable & Description\\
  \midrule
  sex & `male' or  `female'\\
  age & subject age at baseline (in years)\\
  racegrp & race/ethnicity: `black', `hispanic', `white', or `other'\\
  anysub & use of any substance post-detox: `no' or `yes'\\
  cesd & Center for Epidemiologic Studies Depression measure at baseline (high scores indicate more depressive symptoms)\\
  substance & primary substance of abuse: `alcohol', `cocaine', or `heroin'\\
  mcs & SF-36 Mental Component Score (measured at baseline, lower scores indicate worse status)\\
  pcs & SF-36 Physical Component Score (measured at baseline, lower scores indicate worse status)\\
  \bottomrule

  \end{tabular}
\end{table}

\begin{table}
  \caption{Abridged codebook for the `CPS85' dataset}\label{tbl:cps85}

  \begin{tabular}{@{}lp{0.9\textwidth}@{}}
  \toprule
  Variable & Description\\
  \midrule
  wage & wage (US dollars per hour)\\
  educ & number of years of education\\
  race &  `NW' (nonwhite) or `W' (white)\\
  age & age in years\\
  sex & `F' (female) or `M' (male)\\
  married & `Married' or `Single'\\
  exper & number of years of work experience (inferred from age and education)\\
  union & `Union' or `Not Union'\\
  sector & sector of employment: `clerical', `const', `manag', `manuf', `other', `prof', `sales', or `service'\\
  \bottomrule
  \end{tabular}
\end{table}

We recommend starting out with some motivating questions for class discussion:

\begin{itemize}
  \item What does one row of the dataset represent?
 \item Who is included in this dataset? Who is not?
 \item  What kinds of variables are included?
\end{itemize}

After discussing these answers, the students have a better sense of the scope of the dataset, but questions remain regarding the relationships between the variables. The next step is then to show some simple summary statistics and visualizations that help shed light on more targeted questions:
\begin{itemize}
  \item  What are the proportions of men and women in this dataset?
\item What are the proportions of different primary substances of abuse in this dataset?
\item  What does the relationship look like between depression score and overall mental health?
\end{itemize}

We distribute to students a handout (see Appendix A) that contains the code to generate a variety of univariate (histograms and barplots) and bivariate (boxplots and scatterplots) plots. We demonstrate how some of these plots can be drawn in R by typing the corresponding code from the handout into an R Markdown file and compile the results. We also show how easy it is to load the help files for the relevant functions to look at ways of adding third or fourth grouping variables to make multivariate visualizations. We then ask the students to practice statistical thinking by reflecting on what is learned from each plot and discuss as a class. Some plots are not particularly insightful (see Figure~\ref{fig:not_great_plot} for example) relative to others (Figure~\ref{fig:better_plot}). And additional information might be gleaned by including another variable (Figure~\ref{fig:best_plot}). Through these demonstrations, we encourage students to think about what interesting questions might be answered given the data and the tools at their disposal.

\begin{knitrout}
\definecolor{shadecolor}{rgb}{0.969, 0.969, 0.969}\color{fgcolor}\begin{figure}
\includegraphics[width=\maxwidth]{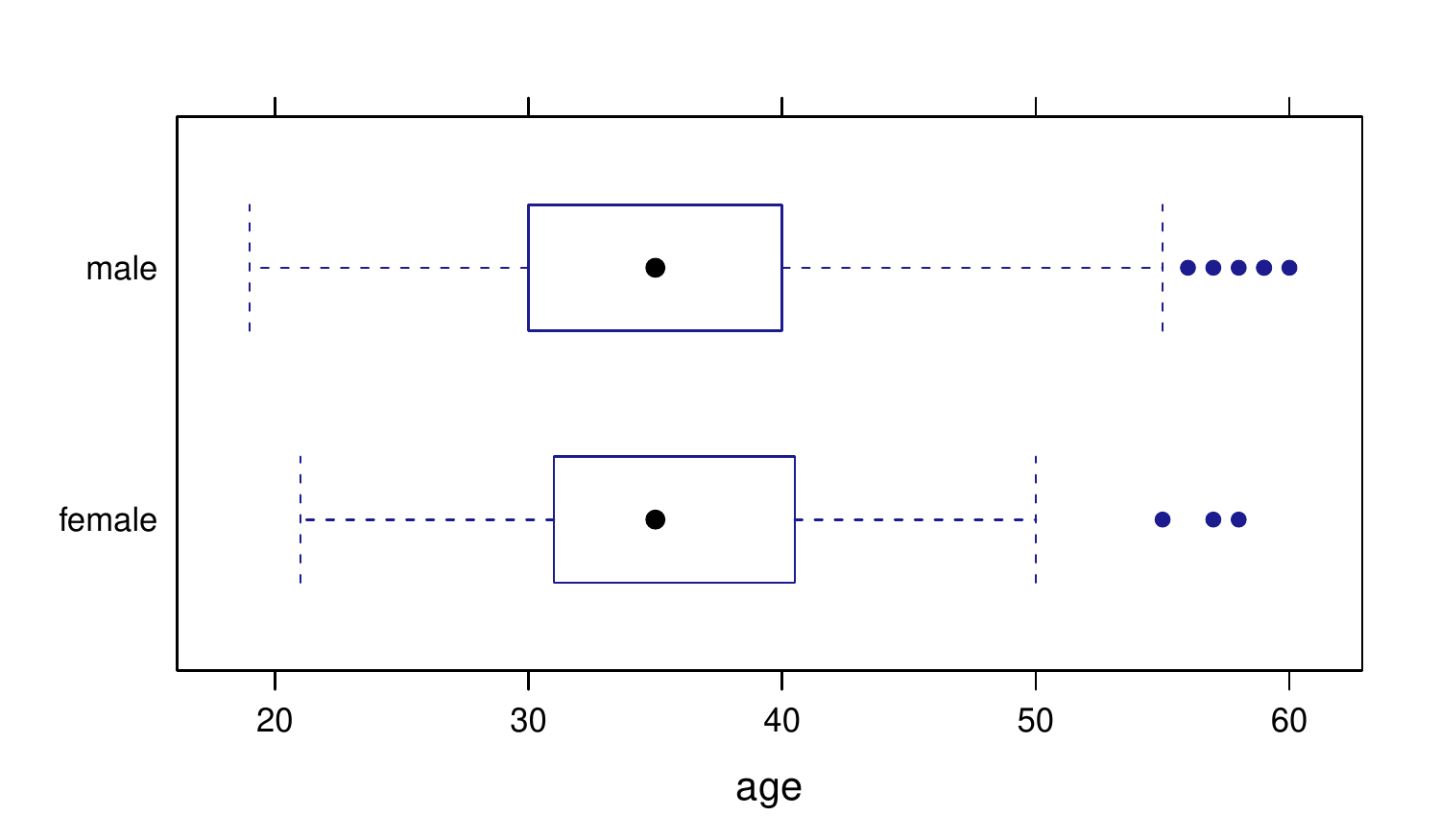} \caption[Boxplot of age by sex of subjects]{Boxplot of age by sex of subjects}\label{fig:not_great_plot}
\end{figure}

\end{knitrout}

\begin{knitrout}
\definecolor{shadecolor}{rgb}{0.969, 0.969, 0.969}\color{fgcolor}\begin{figure}
\includegraphics[width=\maxwidth]{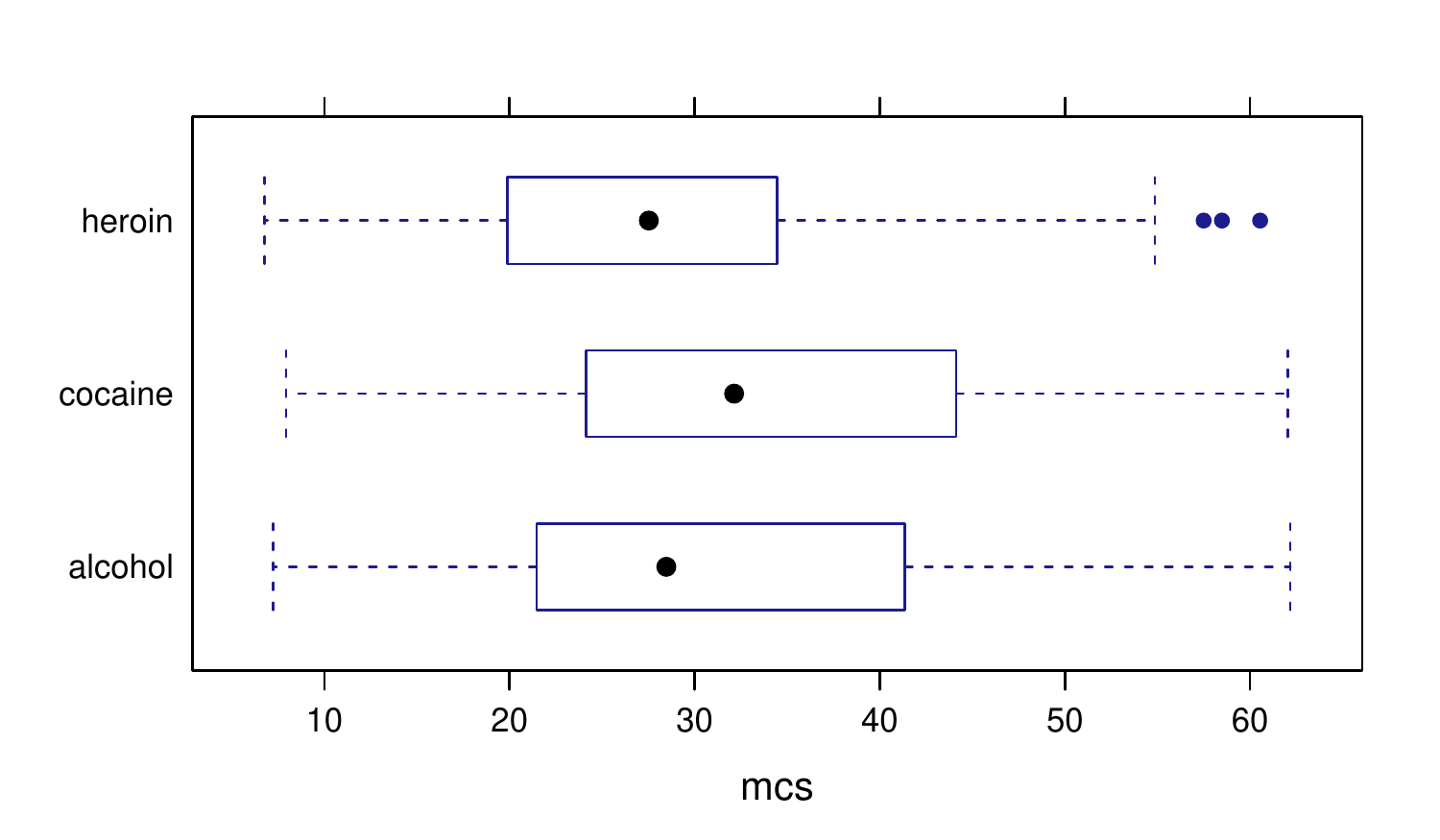} \caption[Boxplot of Mental Component Score by primary substance of abuse]{Boxplot of Mental Component Score by primary substance of abuse}\label{fig:better_plot}
\end{figure}

\end{knitrout}

\begin{knitrout}
\definecolor{shadecolor}{rgb}{0.969, 0.969, 0.969}\color{fgcolor}\begin{figure}
\includegraphics[width=\maxwidth]{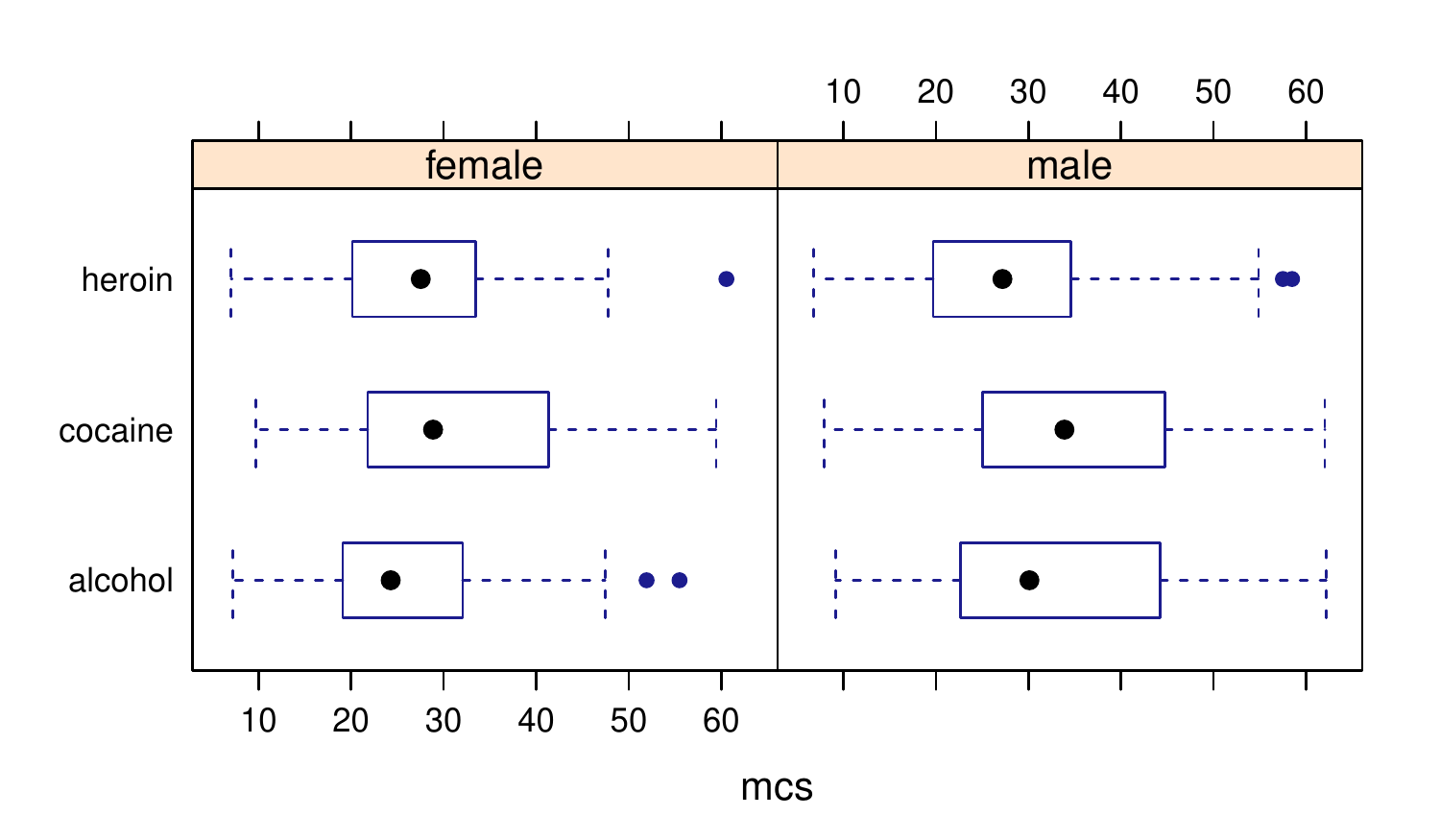} \caption[Boxplot of Mental Component Score by primary substance of abuse, grouped by sex]{Boxplot of Mental Component Score by primary substance of abuse, grouped by sex}\label{fig:best_plot}
\end{figure}

\end{knitrout}

After the brief introduction to R, we then provide students with a different dataset of similar complexity. For this second dataset, we used data from the 1985 Current Population Survey (`CPS85') \citep{berndt1991practice}. Table~\ref{tbl:cps85} summarizes some of the variables contained in this dataset. Note that getting into RStudio and loading the R Markdown template may take some time when students first use R/RStudio and RMarkdown.

The students are provided a single R Markdown worksheet (Appendix B provides an example) in which they are asked to produce three meaningful plots of the new dataset to provide insight on some facet of interest, and write a couple of sentences about each plot to discuss what they learned from it. Students should complete this task in groups of about two or three over a period of 40 to 50 minutes. The deliverable is a compiled R Markdown file, in HTML form, that collates the code, plots, and descriptions.

\subsection{Requirements} 
\label{sec:requirements}
The simplest environment to support this activity is an institutional RStudio server since in this setup, students need only a web browser to run their analyses.
RStudio servers are licensed on a per-user basis but are provided free for academic institutions for teaching purposes. Having an RStudio server that all can access on day one helps to overcome several difficulties: students do not have to download RStudio and R on their own computers, which may present its own sort of problem when installation fails; also, the server can be preloaded with some packages that all can use \citep{ecotsmine}; finally, working from a server eliminates the worry of having insufficient computing power on the students' own computers. Alternatively, the activity could be done within a computer lab where RStudio and necessary packages have been preloaded.

The R Markdown system enables students (and instructors) to break up R code into short, digestable chunks that can then be annotated. The beauty of this system is that allows for the easy creation of presentation-worthy reports showcasing the results of analysis; it seamlessly weaves together R code with plain text to produce a single file containing analysis and explanation.
In this way, R Markdown provides a compelling argument for getting students used to the process of continuously documenting their work and encourages clear presentation of findings \citep{Baumer:2014ud}. With native support within RStudio, and recently bolstered by a new R Notebook feature that provides automatic previews, R Markdown is straightforward for students to work with even on day one of class. Additionally, compiled R Markdown files can be shared publicly via RPubs, a free platform for showcasing R Markdown output\footnote{see \url{https://rpubs.com/about/getting-started}}.

For graphing, we utilize the {\tt mosaic} package \citep{mosaicStudent,Pruim:2015ww} and take advantage of its
simple-to-learn R syntax that helps unite the various different R functions used for data explorations. Conveniently, the syntax for modeling can be described by:

\begin{verbatim}
GOAL(Y ~ X, data= DATASET)
\end{verbatim}
with variations depending on whether there are more (multivariate) or fewer (univariate) variables than the typical outcome and single predictor. Students need to pick a GOAL (e.g. create a scatterplot), specify the variables (X and Y) to study, and the DATASET containing these variables.

As an example, a comparison of mean wages by sex could be generated through the command:
\begin{knitrout}
\definecolor{shadecolor}{rgb}{0.969, 0.969, 0.969}\color{fgcolor}\begin{kframe}
\begin{alltt}
\hlkwd{mean}\hlstd{(wage} \hlopt{~} \hlstd{sex,} \hlkwc{data}\hlstd{=CPS85)} \hlcom{# wage "by" sex}
\end{alltt}
\begin{verbatim}
##    F    M 
## 7.88 9.99
\end{verbatim}
\end{kframe}
\end{knitrout}
while side by side boxplots could be generated by running (see Figure~\ref{fig:bwplot_eg}):
\begin{knitrout}
\definecolor{shadecolor}{rgb}{0.969, 0.969, 0.969}\color{fgcolor}\begin{kframe}
\begin{alltt}
\hlkwd{bwplot}\hlstd{(sex} \hlopt{~} \hlstd{wage,} \hlkwc{data}\hlstd{=CPS85)}
\end{alltt}
\end{kframe}\begin{figure}
\includegraphics[width=\maxwidth]{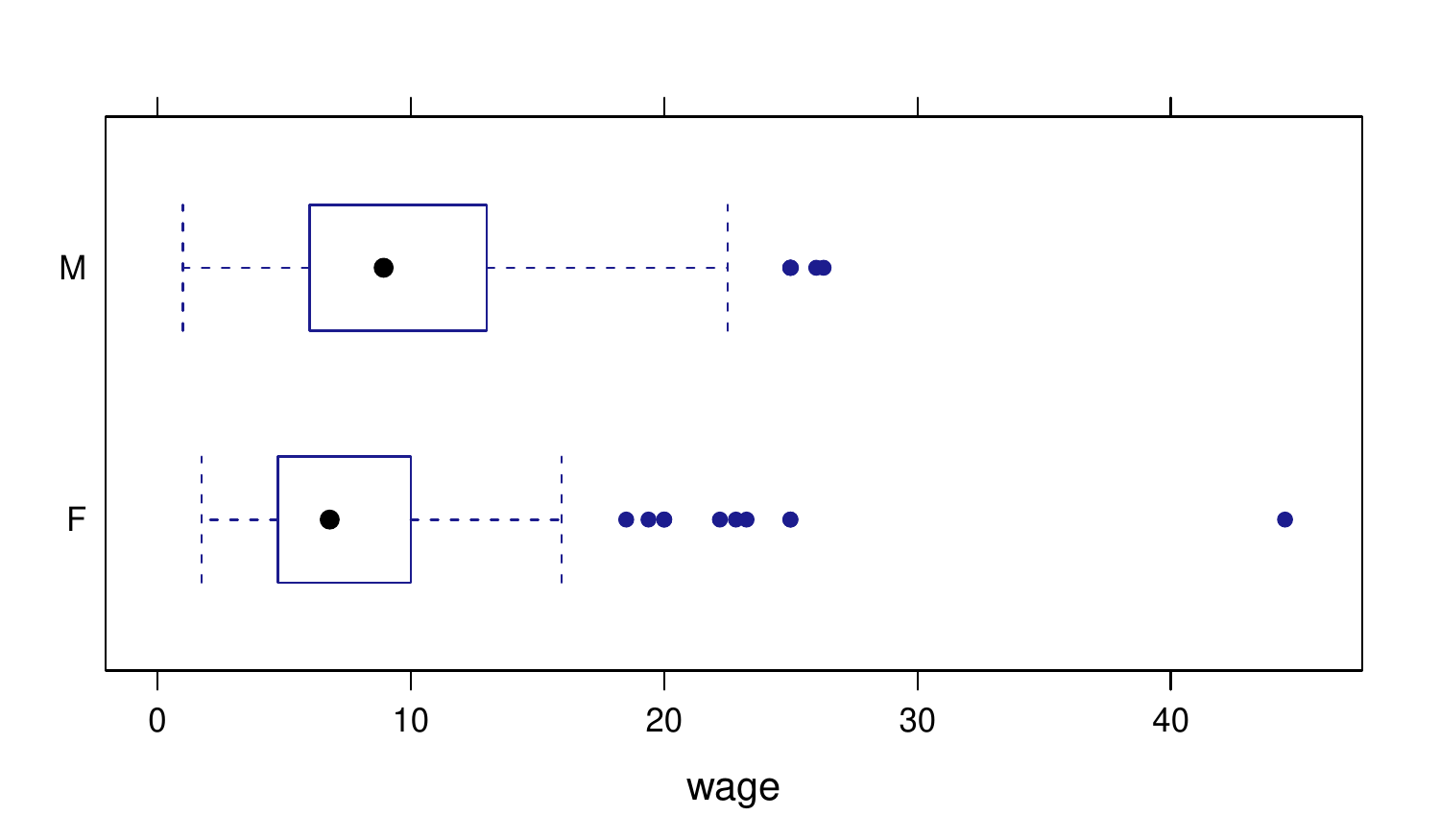} \caption[Boxplot of hourly wage by sex]{Boxplot of hourly wage by sex}\label{fig:bwplot_eg}
\end{figure}

\end{knitrout}
Scatterplots of two quantitative variables can be generated using a similar command (see Figure~\ref{fig:xyplot_eg}):

\begin{knitrout}
\definecolor{shadecolor}{rgb}{0.969, 0.969, 0.969}\color{fgcolor}\begin{kframe}
\begin{alltt}
\hlkwd{xyplot}\hlstd{(wage} \hlopt{~} \hlstd{age,} \hlkwc{data}\hlstd{=CPS85)}
\end{alltt}
\end{kframe}\begin{figure}
\includegraphics[width=\maxwidth]{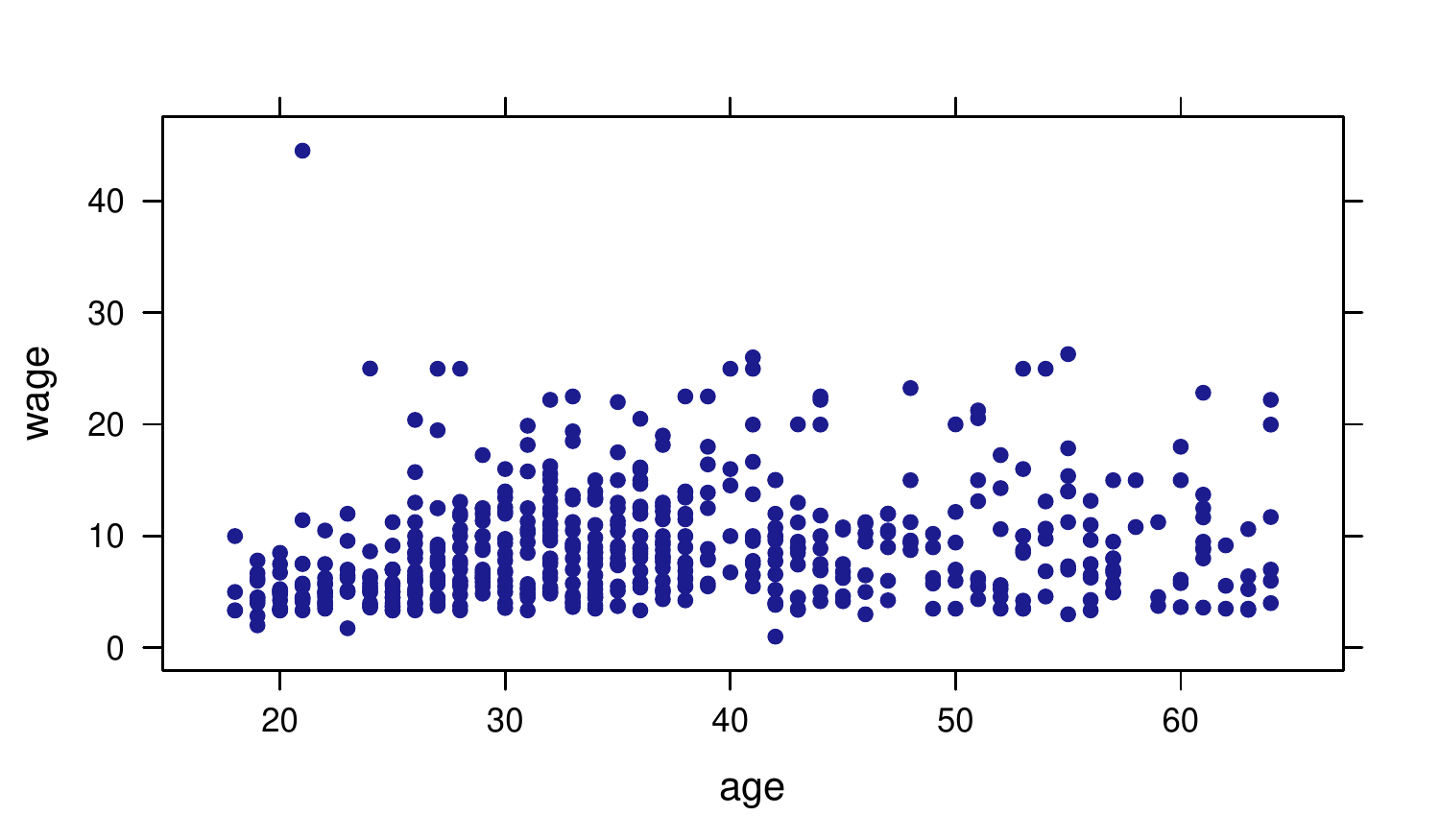} \caption[Scatterplot of hourly wage by age]{Scatterplot of hourly wage by age}\label{fig:xyplot_eg}
\end{figure}

\end{knitrout}

Multivariate displays are straightforward to generate. Figure~\ref{fig:xyplot2_eg} modifies Figure~\ref{fig:xyplot_eg} by adding sex as a grouping variable via the `\texttt{group=}' argument. The `\texttt{auto.key=TRUE}' argument asks for a figure legend matching colors to levels of the grouping variable. The `\texttt{type=}' argument optionally allows multiple geometric layers to be displayed on the same plot; popular options are ``p'' for points and ``r'' for least squares lines.

\begin{knitrout}
\definecolor{shadecolor}{rgb}{0.969, 0.969, 0.969}\color{fgcolor}\begin{kframe}
\begin{alltt}
\hlkwd{xyplot}\hlstd{(wage} \hlopt{~} \hlstd{age,} \hlkwc{group}\hlstd{=sex,} \hlkwc{type}\hlstd{=}\hlkwd{c}\hlstd{(}\hlstr{"p"}\hlstd{,} \hlstr{"r"}\hlstd{),} \hlkwc{auto.key}\hlstd{=}\hlnum{TRUE}\hlstd{,} \hlkwc{data}\hlstd{=CPS85)}
\end{alltt}
\end{kframe}\begin{figure}
\includegraphics[width=\maxwidth]{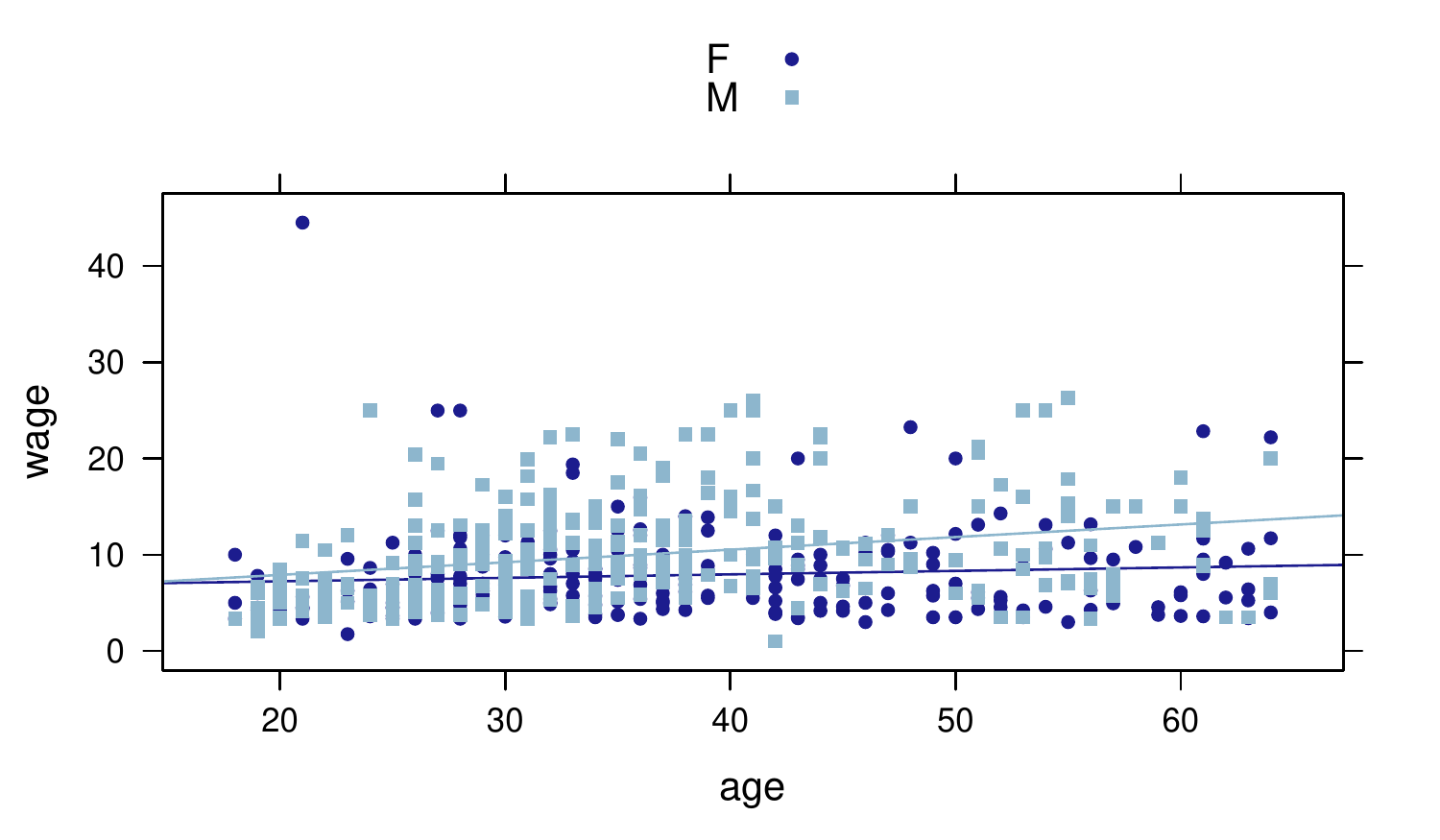} \caption[Scatterplot of hourly wage by age, grouped by sex]{Scatterplot of hourly wage by age, grouped by sex}\label{fig:xyplot2_eg}
\end{figure}

\end{knitrout}

\newpage
If we want to incorporate a grouping variable using facetting as opposed to using multiple colors (such as in Figure~\ref{fig:best_plot}), we could use a vertical pipe to specify our grouping variable. Figure~\ref{fig:best_plot} was produced using the command:

\begin{knitrout}
\definecolor{shadecolor}{rgb}{0.969, 0.969, 0.969}\color{fgcolor}\begin{kframe}
\begin{alltt}
\hlkwd{bwplot}\hlstd{(substance} \hlopt{~} \hlstd{mcs} \hlopt{|} \hlstd{sex,} \hlkwc{data}\hlstd{=HELPrct)}
\end{alltt}
\end{kframe}
\end{knitrout}

We also recommend making use of the \texttt{mosaicData} R package \citep{mosaicData}, which contains a number of ready-to-use datasets that would serve well for this activity. Both of our selected datasets are contained in this package.

We have used this lab activity in our classes with an average of 25 students. However, we believe that if course assistants are available to maintain the ratio of one assistant per 15 to 20 students, this activity can scale easily for classes with more students. The purpose of course assistants is to be able to help with hurdles as they arise -- missing parentheses, misspelled variables/functions, incorrect capitalization -- all of which are commonly experienced by a student working in R for the first time. We detail a number of other common pitfalls in the Discussion.

\subsection{Selecting Datasets} 
\label{sub:selecting_datasets}
It is important to use datasets that are of interest to students \citep{gould2010statistics}. The datasets we have selected contain a good mix of demographic data (like age, race, and years of education), which students easily comprehend, along with some other aspect that is relevant to them (depression scores and alcohol usage in the `HELPrct' dataset and wages, work experience, and sector in the `CPS85' dataset). In addition to the two datasets presented here, we also recommend the 2013 New York City flight delays dataset \citep{nycflights13}, the IMDB movie ratings dataset \citep{imdb}, and the 2006 New Haven residential property dataset \citep{barcode}, accessible in R packages \texttt{nycflights13}, \texttt{ggplot2movies}, and \texttt{barcode}, respectively. We should note that larger datasets, say, with sample size larger than 1000, could be tricky for students to work with due to the time it takes to plot a large number of points and excessive overplotting.

Having about 8 to 15 variables provides a good variety of different possibilities for exploration. Too few variables may yield a smaller chance to identify questions of interest and too many variables would overwhelm. A good mix of categorical and quantitative variables would provide more opportunities to practice with different kinds of data summaries and visualizations. Needless to say, having a large number of rows in the dataset is also important so that, for example, if a student wanted to include explorations of two variables by subgroups of a third, the sample sizes would still be substantial enough to render interesting visualizations.

We also recommend selecting datasets that are already built into R or an R package (or include code to download the data directly into the R Markdown file using the \texttt{read.csv()} function in conjunction with a weblink). The benefit of using a dataset that can be loaded with minimal effort is that we can head straight into exploratory data analysis. An added bonus is that most built-in datasets are described using a help page that provides a codebook and description of the variables. The ability to access and interpret help files is an important skill that will inevitably become useful as students delve into the deeper levels of R computing as time progresses.

\section{RESULTS} 
\label{sec:results}
The authors collected student feedback concerning this activity from introductory statistics students at Amherst College. We received permission to report anonymous student feedback and to show examples of student work from those groups of students who have provided consent (Amherst College Institutional Review Board approval \#15-028). Out of 50 total students, 39 completed an electronic consent form affirming that their work may be shared, while the other 11 did not complete the form. Rather than pick a few of the examples to share in this article, we have posted all of them online\footnote{\url{http://xiaofei-wang.com/research/vislab/}}.

The students noted strengths and weaknesses of the activity when queried at the one month mark of the course. The following positive aspects of the activity were excerpted from the survey:

\begin{itemize}
\item The activity got us to immediately start working in R in a hands-on way. We started with a brief overview and then got to experience the eccentricities of coding in R. Eventually, I became more comfortable with working with R.
\item I got to work with data.
\item I got to meet and work with my classmates. I appreciate that the activity was self-guided and that it gave opportunity for teamwork.
\item The activity was challenging and gave us an accurate depiction of how the class is run. We were able to ask for help from the instructor and the lab assistant when we got stuck.
\end{itemize}

Below are areas where students noted room for improvement:

\begin{itemize}
  \item The pace was a bit fast. A bit more direction would improve the experience.
 \item It would be beneficial to discuss the plots (histograms, bargraphs, boxplots, etc.) and compare and contrast them. They utilize different types of variables; I realized this after forming an incorrect plot.
  \item You asked us to write descriptions before we really knew how to describe.

\end{itemize}

In summary, most students acknowledged the importance of R and appreciated the chance to experience it early on with support from the instructor and lab assistant. Moreover, the activity served as an ice breaker between classmates, facilitating subsequent group-based activities.

At the same time, not all students appreciated the whirlwind tour of R offered by this activity. Some students craved more guidance and a slower pace. Some students expressed that they felt lost through the exercise. Indeed, the group work began after only fifteen minutes of instruction. However, we believe it is acceptable for a first-week activity to leave students with questions unanswered and a desire to learn more. We recommend setting the right expectations by conveying to students that the activity is intended to provide a first exposure to R, with mastery to be achieved later in the semester.


At the end of the class period, students were able to code and generate a presentable HTML file containing three plots and summaries using a reproducible analysis framework. Incorporating this experience into the first week of class dispelled the sense that the early classes would simply be a rehash of mean, median, and mode, and provided additional confidence for subsequent interactions with R.

Given the critical feedback, a follow-up class discussion might be inserted  after the students have learned more about data visualization through readings and in-class examples. This discussion might begin with a critique of a few lab submissions from the activity. For example, we might find an instance where a student plotted a histogram but called it a bargraph in the accompanying description or vice versa. Upon revisiting the activity, some students will have learned that bargraphs are meant for categorical variables and histograms are meant for quantitative variables; reviewing the mistakes from the lab activity helps reinforce these new concepts. We might also take the opportunity to critique the phrasing of some of the descriptions that were written. A student's description of a graph might say ``this is a histogram that depicts the distribution of $x$'', to which we can now agree as a class that a better insight would discuss ``surprisingly, we see that $x$ is actually bimodal, with peaks at 3 and 5.'' This exercise gives students a second look to reflect upon their previous work and see how far they have come since their initial foray.

\section{DISCUSSION} 
\label{sec:discussion}

An important learning outcome in any statistics class is for students to begin to think like a statistician. Specifically, we believe this consists of repeated practice with posing statistical questions and answering them with evidence backed by data. Software makes this an achievable practice, even in the first week of class, if we exploit the students' curiosity about the world around them. If we provide a dataset about which students have some contextual understanding (even better, misunderstanding), they will naturally pose interesting questions.

In our experience, there are some common pitfalls that students encounter during this activity. When students first download the template R Markdown file, some browsers will force a .txt extension on the file. Students have to change the extension back to .Rmd in order to proceed. One way around this issue is to simply copy and paste the contents that appear as raw text in their browser window into a brand new R Markdown file. Sometimes this act of copying and pasting introduces leading whitespace before some code chunks, which have to be manually deleted in order for the file to compile. Some students may try a lot of different plots in the RStudio console before picking their favorites to submit. In the process of copying their work from the console to their R Markdown script, they may include the ``\texttt{+}'' and ``\texttt{>}'' signs that then break the compilation process. Some students have difficulty distinguishing between code chunks from regular text. In several instances, students learned that the ``\texttt{\#}'' symbol creates a comment in R, but placed these comments outside of a code chunk, in which case the comment gets printed as top-level header font in Markdown. Furthermore, some students do not realize that a button or keyboard shortcut allows them to create code chunks, so instead they manually type in the code chunk headers and footers. With incorrect syntax, they then run into compiling issues. All of these pitfalls are part of the learning curve; we expect that students will run into a number of these issues sooner or later when learning R. Experiencing these issues in class gives students immediate assistance when they do arise and helps minimize the friction of learning new software. To make the activity run as smoothly as possible, we highly recommend having one course assistant per 15 to 20 students during this activity.

To reiterate, our proposed activity does not aim to produce experts at data visualization or R coding; rather, it is intended to serve as a pedagogical tool to inspire multivariate thinking. Our goal is to motivate students to ask good, statistical questions and then attempt to answer them with data and a minimal amount of computing.
Being able to shed light on those questions, albeit without the rigor of considering significance, within a class period in the first week of class is extremely empowering and helps to whet their appetite for more to come.

The activity is extensible depending on how much time can be allotted for it. On some occasions, we have asked students to share their compiled HTML files on RPubs. This approach is attractive since it allows student findings to be shared with the class as a whole by sharing the appropriate RPubs link. In the instances where we added this step, we found that students took more time to polish their work, taking greater ownership in the final published product. If additional time is available (perhaps in a second class period), some groups of students can present their plots to the rest of their class. This helps to develop communication skills, overcome the fear of speaking to ones' classmates, and share insights, all early in the course.

\newpage

\bibliography{dataviz}

\newpage
\section*{Appendix A: Lab Handout} 
\label{sec:appendix_a_handout_for_activity}
\small
\subsection*{The Template}
The template for most functions (from the \texttt{mosaic} package in R) is:
\begin{center}
\verb|goal(     ~    , data =     )|
\end{center}
\subsection*{Getting R to Work}
Each command you type should be guided by the following 2 questions:
\begin{enumerate}
  \item What do you want R to do?
  \item What must R know to do that?
\end{enumerate}
\subsection*{Exploring the Data}
In this course, we'll work with datasets that have a combination of quantitative and categorical variables. Oftentimes, an important first step (before doing any analysis) is to explore the data. Here are some plots that are frequently used to visually display the data.

\subsubsection*{Univariate Summaries}

\begin{multicols}{2}
\begin{knitrout}
\definecolor{shadecolor}{rgb}{0.969, 0.969, 0.969}\color{fgcolor}\begin{kframe}
\begin{alltt}
  \hlkwd{tally}\hlstd{(}\hlopt{~} \hlstd{sex,} \hlkwc{data}\hlstd{=HELPrct)}
\end{alltt}
\begin{verbatim}
## 
## female   male 
##    107    346
\end{verbatim}
\begin{alltt}
  \hlkwd{bargraph}\hlstd{(}\hlopt{~} \hlstd{sex,} \hlkwc{data}\hlstd{=HELPrct)}
\end{alltt}
\end{kframe}
\includegraphics[width=\maxwidth]{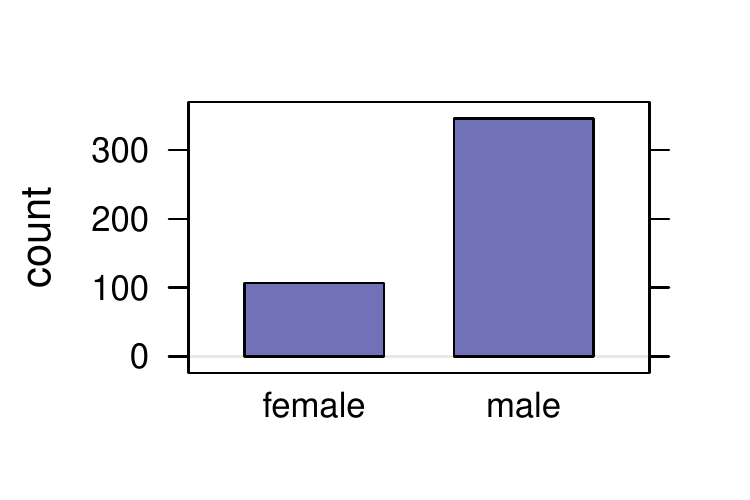} 

\end{knitrout}

\columnbreak

\begin{knitrout}
\definecolor{shadecolor}{rgb}{0.969, 0.969, 0.969}\color{fgcolor}\begin{kframe}
\begin{alltt}
  \hlkwd{favstats}\hlstd{(}\hlopt{~} \hlstd{age,} \hlkwc{data}\hlstd{=HELPrct)}
\end{alltt}
\begin{verbatim}
##  min Q1 median Q3 max mean   sd   n missing
##   19 30     35 40  60 35.7 7.71 453       0
\end{verbatim}
\begin{alltt}
  \hlkwd{histogram}\hlstd{(}\hlopt{~} \hlstd{age,} \hlkwc{data}\hlstd{=HELPrct)}
\end{alltt}
\end{kframe}
\includegraphics[width=\maxwidth]{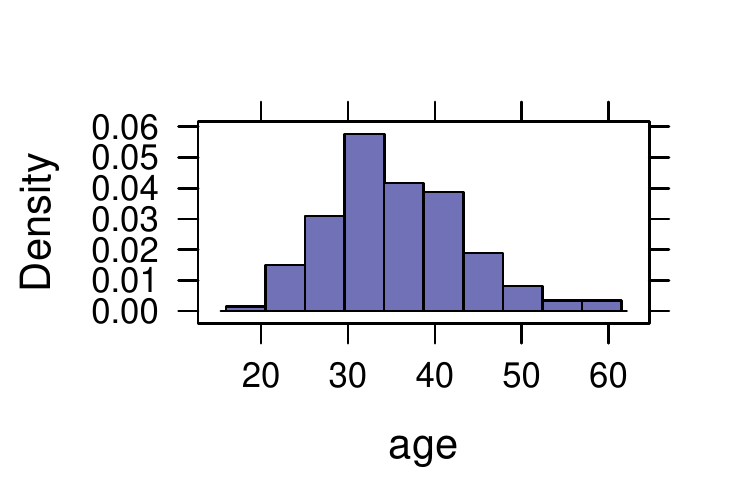} 

\end{knitrout}
\end{multicols}
\newpage
\subsubsection*{Bivariate Summaries}

\begin{multicols}{2}
Categorical var. vs. categorical var.
\begin{knitrout}
\definecolor{shadecolor}{rgb}{0.969, 0.969, 0.969}\color{fgcolor}\begin{kframe}
\begin{alltt}
  \hlkwd{tally}\hlstd{(homeless} \hlopt{~} \hlstd{sex,} \hlkwc{data}\hlstd{=HELPrct)}
  \hlkwd{bargraph}\hlstd{(}\hlopt{~} \hlstd{sex,} \hlkwc{group} \hlstd{= homeless,}
                  \hlkwc{data}\hlstd{=HELPrct,}
                  \hlkwc{auto.key}\hlstd{=}\hlnum{TRUE}\hlstd{)}
\end{alltt}
\end{kframe}
\end{knitrout}
\begin{knitrout}
\definecolor{shadecolor}{rgb}{0.969, 0.969, 0.969}\color{fgcolor}\begin{kframe}
\begin{verbatim}
##           sex
## homeless   female male
##   homeless     40  169
##   housed       67  177
\end{verbatim}
\end{kframe}
\includegraphics[width=\maxwidth]{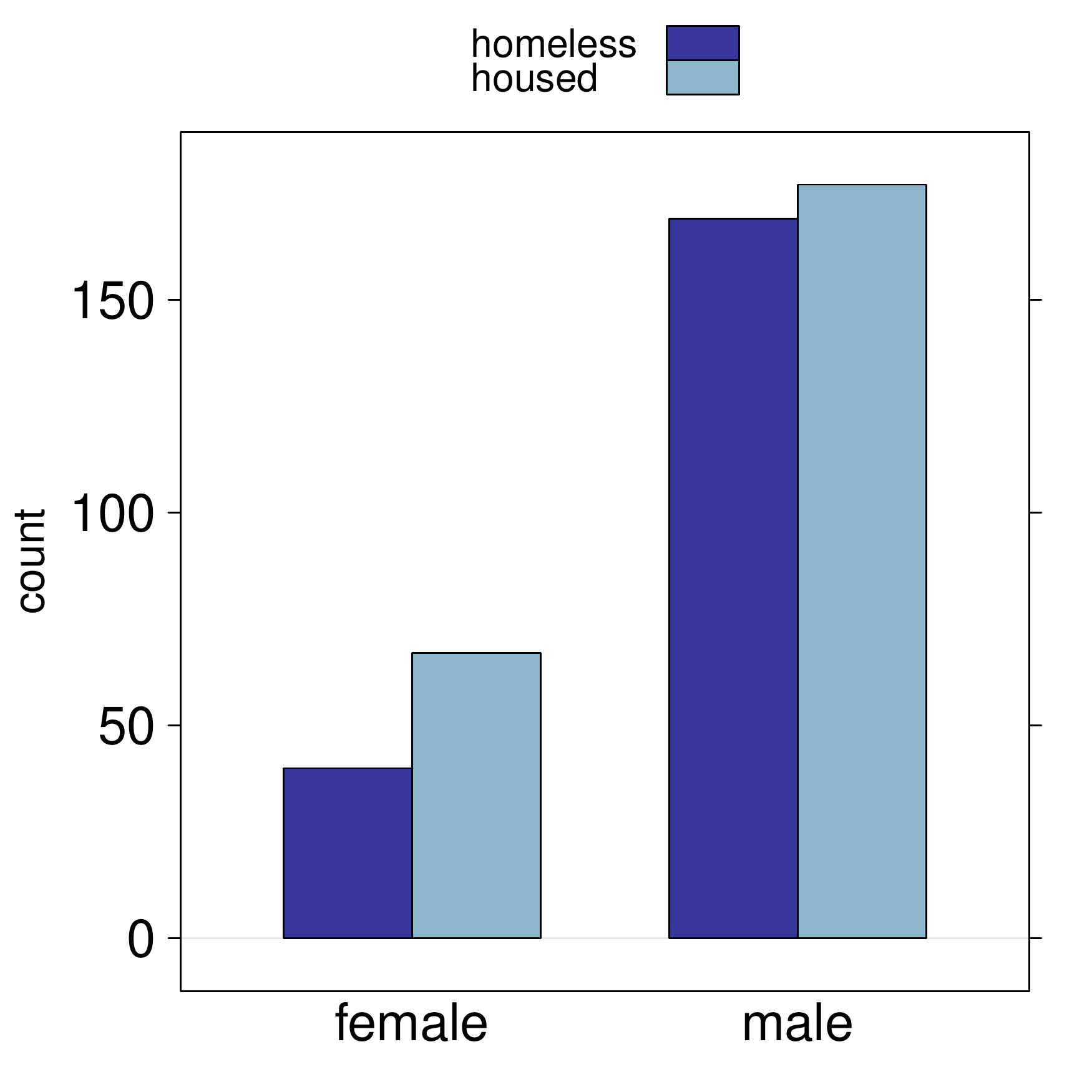} 

\end{knitrout}
\columnbreak
Quantitative var. vs. quantitative var.
\begin{knitrout}
\definecolor{shadecolor}{rgb}{0.969, 0.969, 0.969}\color{fgcolor}\begin{kframe}
\begin{alltt}
  \hlkwd{cor}\hlstd{(i1} \hlopt{~} \hlstd{age,} \hlkwc{data}\hlstd{=HELPrct)}
  \hlkwd{xyplot}\hlstd{(i1} \hlopt{~} \hlstd{age,} \hlkwc{data}\hlstd{=HELPrct)}
\end{alltt}
\end{kframe}
\end{knitrout}
\begin{knitrout}
\definecolor{shadecolor}{rgb}{0.969, 0.969, 0.969}\color{fgcolor}\begin{kframe}
\begin{verbatim}
## [1] 0.207
\end{verbatim}
\end{kframe}
\includegraphics[width=\maxwidth]{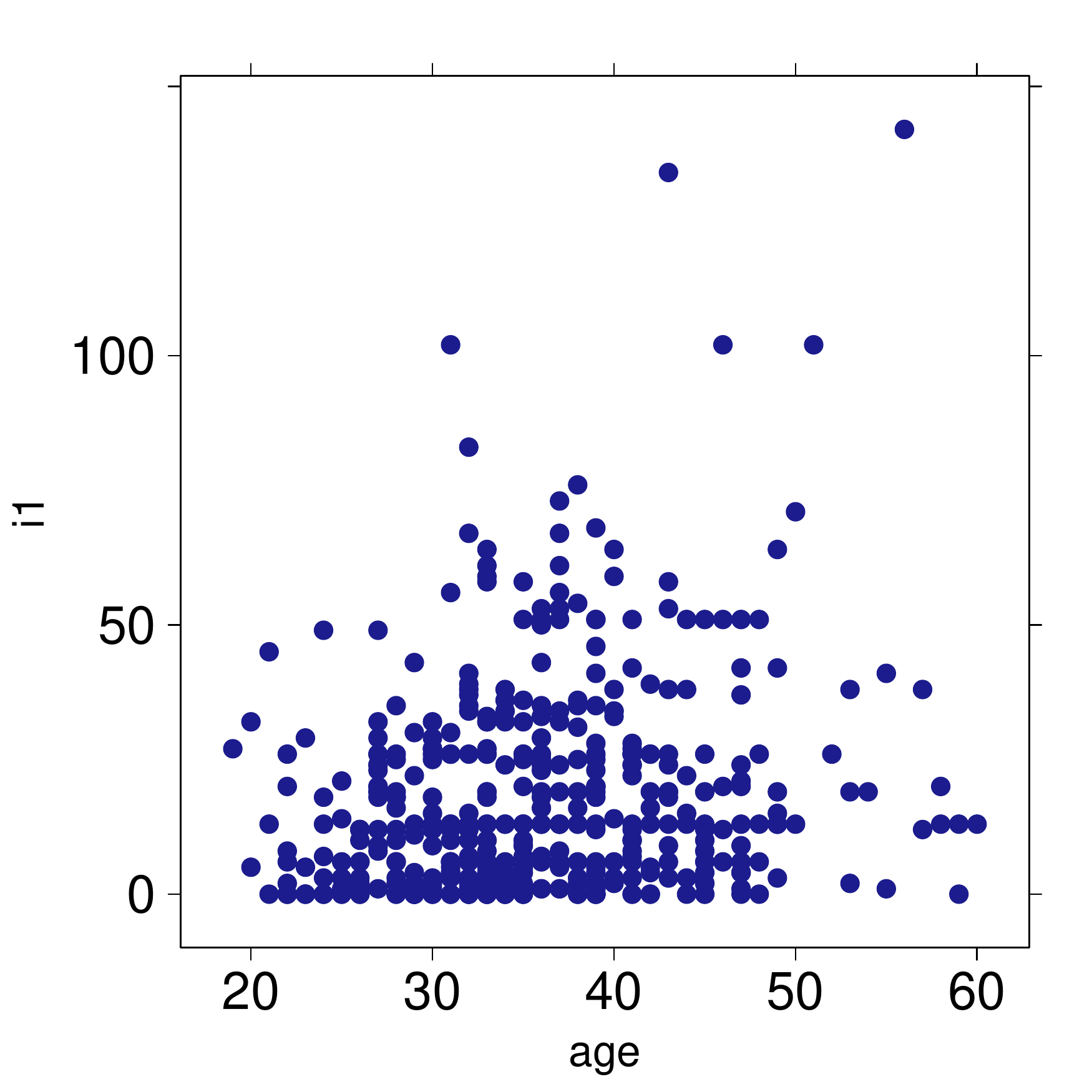} 

\end{knitrout}
\end{multicols}
\newpage
Categorical var. vs. quantitative var.
\begin{knitrout}
\definecolor{shadecolor}{rgb}{0.969, 0.969, 0.969}\color{fgcolor}\begin{kframe}
\begin{alltt}
  \hlkwd{favstats}\hlstd{(age} \hlopt{~} \hlstd{racegrp,} \hlkwc{data}\hlstd{=HELPrct)}
  \hlkwd{bwplot}\hlstd{(age} \hlopt{~} \hlstd{racegrp,} \hlkwc{data}\hlstd{=HELPrct)}
\end{alltt}
\end{kframe}
\end{knitrout}
\begin{knitrout}
\definecolor{shadecolor}{rgb}{0.969, 0.969, 0.969}\color{fgcolor}\begin{kframe}
\begin{verbatim}
##    racegrp min   Q1 median   Q3 max mean   sd   n missing
## 1    black  20 31.0     35 39.0  60 35.7 7.08 211       0
## 2 hispanic  21 28.2     32 36.2  55 33.2 7.99  50       0
## 3    other  22 30.0     34 40.5  48 35.0 7.66  26       0
## 4    white  19 30.0     36 42.0  58 36.5 8.28 166       0
\end{verbatim}
\end{kframe}
\includegraphics[width=0.5\textwidth]{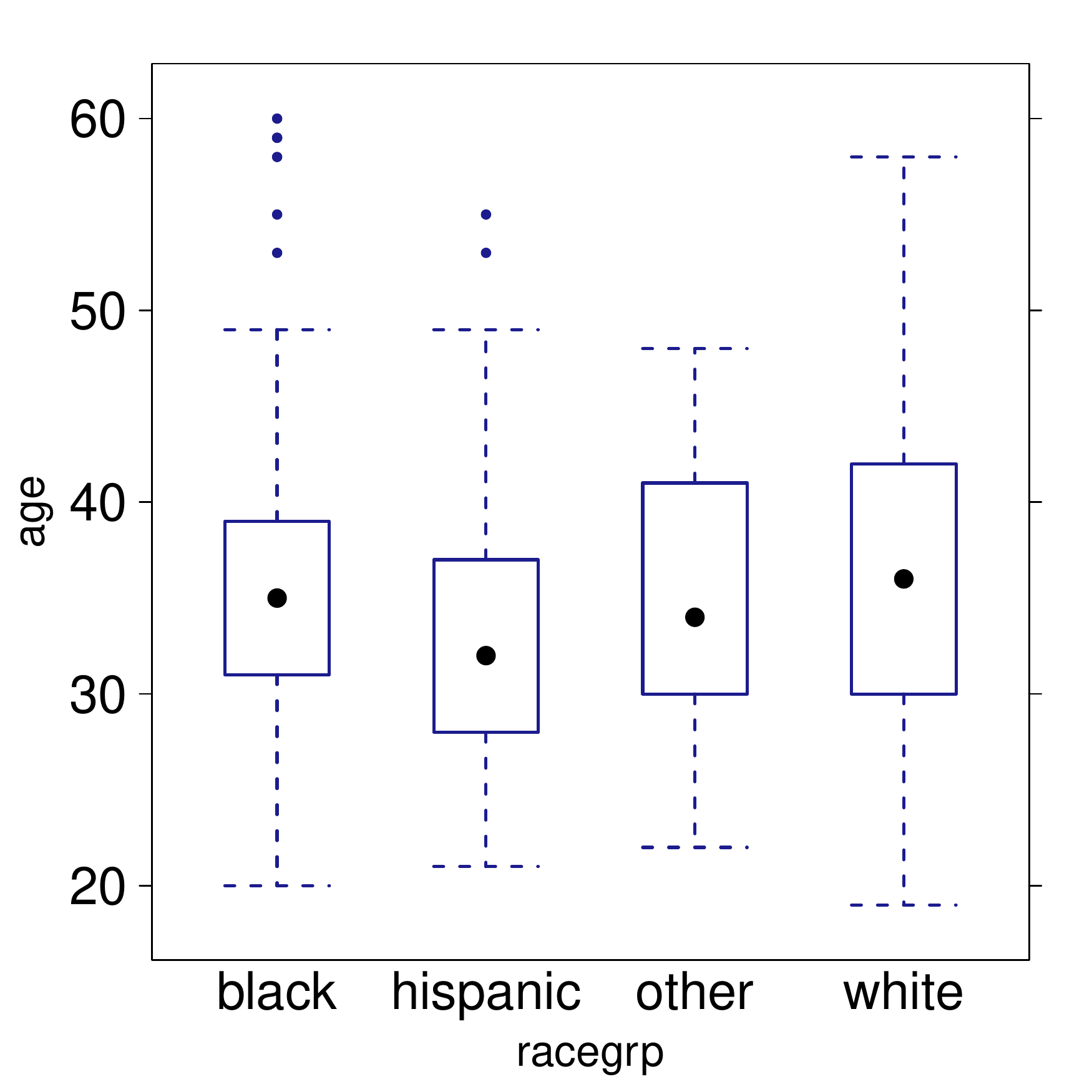} 

\end{knitrout}
\subsection*{Helpful Tips}
\begin{itemize}
\item R is case sensitive: \verb|x| is not the same thing as \verb|X|.
\item In the console, \verb|> | means R is ready for a new command, whereas \verb|+ | means R is \textit{waiting for you to finish} an existing command. Hitting ESC gets you out of the latter scenario if you're there by accident.
\item Not sure what a function like \verb|msummary()| does? Type the function name preceded by a question mark, like this: \verb|?msummary| to get help. Scroll down to Examples -- replicate some of these on your own.
\item If R throws you an error, read it before you panic. Usually, the error is more interpretable than you think!
\end{itemize}

\newpage
\section*{Appendix B: Lab Activity} 
\label{sec:appendix_b_lab_activity}
\subsection*{Instructions}\label{instructions}
\large
Please delete this entire section before you
submit your file to RPubs!

In your groups, explore the \texttt{CPS85} dataset within the
\texttt{mosaicData} package to try to find some interesting insights. You may want to type \texttt{?CPS85} and
\texttt{head(CPS85)} to get a glimpse at what this dataset contains.
Next, start exploring the dataset using plots, tables, and other numeric
summaries. Select 3 favorite plots and tell a story (in writing) about
each of them. Extra brownie points if you can weave the 3 plots together
into one cohesive story.

\subsection*{PLOT 1}\label{plot-1}

\begin{knitrout}
\definecolor{shadecolor}{rgb}{0.969, 0.969, 0.969}\color{fgcolor}\begin{kframe}
\begin{alltt}
\hlcom{# put the code for your plot here}
\end{alltt}
\end{kframe}
\end{knitrout}

(Include the description for your plot here.)

\subsection*{PLOT 2}\label{plot-2}

\begin{knitrout}
\definecolor{shadecolor}{rgb}{0.969, 0.969, 0.969}\color{fgcolor}\begin{kframe}
\begin{alltt}
\hlcom{# put the code for your plot here}
\end{alltt}
\end{kframe}
\end{knitrout}

(Include the description for your plot here.)

\subsection*{PLOT 3}\label{plot-3}

\begin{knitrout}
\definecolor{shadecolor}{rgb}{0.969, 0.969, 0.969}\color{fgcolor}\begin{kframe}
\begin{alltt}
\hlcom{# put the code for your plot here}
\end{alltt}
\end{kframe}
\end{knitrout}

(Include the description for your plot here.)

\end{document}